\documentclass{ws-procs9x6}

\begin{document}

\title{Monte-Carlo-Simulations of Stochastic Differential Equations\\ 
at the Example of the\\
Forced Burgers' Equation}

\author{Dirk Homeier}

\address{Institut f\"{u}r theoretische Physik, Westf\"{a}lische Wilhelms-Universit\"{a}t,\\
M\"{u}nster, Germany\\
E-mail: homeierd@uni-muenster.de\\
http://pauli.uni-muenster.de}

\author{Karl Jansen$^{1*}$, David Mesterhazy$^2$, Carsten Urbach$^2$ }

\address{$^1$DESY Zeuthen, Germany\\
$^2$Humboldt-Universit\"{a}t Berlin, Germany\\
E-mail: $^*$karl.jansen@desy.de}

\begin{abstract}
We investigate the behaviour of stochastic differential equations, especially Burgers'
eq., by means of Monte-Carlo-techniques. By analysis of the produced configurations, we show
that direct and often intuitive insight into the fundamentals of the solutions to the underlying
equation, like shock wave formation, intermittency and chaotic dynamics, can be obtained. We
also demonstrate that very natural constraints for the lattice parameters are sufficient 
to ensure stable calculations for unlimited numbers of Monte-Carlo-steps.
\end{abstract}


\bodymatter
\section{Introduction and Motivation}
Hydrodynamic turbulence remains a basically unsolved problem of modern physics. This is especially
noticeable as the fundamentals seem to be fairly easy --- the Navier-Stokes-equations for the
velocity and pressure fields $v$ and $p$,
\begin{equation}
    \partial_{t}v_{\alpha} + v_{\beta} \partial_{\beta}v_{\alpha} - \nu \nabla^{2} v_{\alpha}
  + \frac{1}{\rho}\partial_{\alpha}p = 0,
\end{equation}
 with the additional constraint
\begin{equation}  \partial_{\alpha}v_{\alpha} = 0,
\end{equation} express
the conservation of momentum in a classical, newtonian, incompressible fluid of viscosity $\nu$
and density $\rho$. Laminar flows are reproduced very accurately; in the turbulent regime, it still
is an open question how the universal characteristics of a flow, the scaling exponents $\xi_p$ of
the structure functions $S_p$ of order $p$, defined by 
\begin{equation}
S_p(x) := \left\langle [ |v(r+x) - v(r)) |^{p}\right\rangle_r \propto |x|^{\xi_p},
\end{equation}
can be extracted from the basic equations. Intermittency is reflected by exponents $\xi_p$ that
differ from those expected by dimensional analysis.\\
Monte-Carlo-simulations
enable us to analyze turbulent flow patterns in detail, to gain direct insight into the formation
of localized structures and their behaviour, and to measure observables like the scaling 
exponents. \\
Instead of working with the full Navier-Stokes-eqs.,
we decided to elaborate the methods using Burgers' eq.
\begin{equation}
    \partial_{t}v_{\alpha} + v_{\beta} \partial_{\beta}v_{\alpha} - \nu \nabla^{2} v_{\alpha}
= 0, \label{dh:eqburg}
\end{equation}
which
can be interpreted as the flow equation for a fully compressible fluid.
A finite viscosity $\nu$ and energy dissipation $\epsilon$
provide a dissipation length scale $\lambda$
corresponding to the Kolmogorov-scale in Navier-Stokes-turbulence:
\begin{equation}\lambda = \left( \frac{\nu^{3}}{\epsilon}\right)^{\frac{1}{4}}.\end{equation}
 Besides being of
interest of its own (e.g. in cosmology), working with eq. (\ref{dh:eqburg}) has a number of 
technical advantages:\footnote{For a more complete overview, see \cite{dh:bibbec} .}
\begin{itemize}
\item{Burgers' eq. is local, while the incompressibility condition acts as a nonlocal
interaction in Navier-Stokes-turbulence.}
\item{The fundamental solutions to Burgers' eq. are well-known; in the limit of vanishing 
viscosity (Hopf-eq.), these form singular shocks as seen in fig. (\ref{dh:fig1}). The 
dissipation-scale provides an UV-regularization of the shock structures.
\begin{figure}
\begin{center}
\includegraphics[width=2.25in]{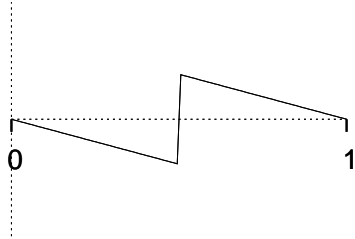}
\caption{Typical solution of Burgers' eq. in the limit of vanishing viscosity. The graph shows
$v(x, t)$ as a function of $x$ in periodic boundary conditions at constant $t$.}
\label{dh:fig1}
\end{center}
\end{figure}}
\item{A huge variety of analytical methods have been applied to Burgers eq., giving results that
can directly be compared to numerical measurements; and the origin of intermittency is well 
understood.}
\end{itemize}
\section{Path Integral}
For the moment, we restrict our work to 1-dimensional Burgulence which already shows intermittent
statistics. A random force $f$ has to be introduced into eq. (\ref{dh:eqburg}) so that solutions 
are statistically homogenous in time:
\begin{equation}
    \mathbf{B}[v]\doteq \partial_{t}v + v \frac{\partial}{\partial x}v - \nu 
\frac{\partial^2}{\partial x^2} v
= f,
\end{equation}
where we model the stochastic force to be gaussian distributed with zero mean, energy injection 
rate $\epsilon$ and correlation length $\Lambda$:
\begin{eqnarray}
\langle f(x,t)\rangle& =& 0, \\
\chi^{-1}(x, x'; t,t') \doteq \langle f(x, t)f(x', t') \rangle &=& \epsilon \delta(t-t')\exp\left(
-\frac{|x-x'|}{\Lambda}\right).
\end{eqnarray}
We then expect intermittent
statistics to be found within the inertial subrange $\lambda \ll x \ll \Lambda$.
We can further identify a characteristic velocity at the injection scale, $u_0 = 
(\epsilon \Lambda)^{1/3}$,
and a Reynolds-number 
\begin{equation} \label{dh:egre}
\mathit{Re} = (\epsilon \Lambda^4/\nu^3)^{1/3}.
\end{equation}
The path integral is introduced via the standard Martin-Siggia-Rose-formalism, giving for the
generating functional\footnote{The functional determinant can be shown not to contribute for
local theories, see e.g. \cite{dh:bibhmpv} .}
\begin{eqnarray}
Z[J] &=& \int \mathcal{D}v \, \mathcal{D}f \, \delta(\mathbf{B}[v] - f) \exp \left(-\frac{1}{2}
\int f \chi f +\int Jv\right) \\
&=&\int \mathcal{D}v \, e^{-S[v; J]},
\end{eqnarray}
with the action
\begin{equation}
S[v; J] = \int dx dt dx' dt' \, \left(\mathbf{B}[v(x,t)]\chi(x,x'; t,t') \mathbf{B}[v(x',t')] + 
J(x,t)v(x,t)\right).
\end{equation}
Beginning from an equivalent path integral, it has been shown that intermittent statistics of 
Burgers' eq. can be understood in terms of instanton solutions (\cite{dh:bibbfkl}).
\section{Monte-Carlo-Simulations}
We discretized the above action onto a rectangular lattice with $L$ sites in space-, and $T$ sites
in time-direction. Derivatives have 
been written in a symmetric (Stratanovich-) prescription. Lattice spacings will be denoted 
$\Delta x$ and $\Delta t$, respectively. We mainly used a heat bath algorithm on single
nodes.
\subsection{Lattice parameters}
To indentify the lattice parameters with the constants of the continuum theory, we first notice
that the viscosity has to be defined as\footnote{For example, this can be shown via the continuum
limit of the symmetric random walk, leading to the diffusion eq..}
\begin{equation}
\nu = \alpha \frac{(\Delta x)^2}{\Delta t},
\end{equation}
in which we define the arbitrary constant $\alpha = 1$. The so-defined $\nu$ gives us $\mathit{Re}$
according to eq. (\ref{dh:egre}). We further find that the dissipation length $\lambda$ is related 
to the correlation length $\Lambda$ simply by
\begin{equation}
\lambda = \frac{\Lambda}{\mathit{Re}^{3/4}}.
\end{equation}
In any practical application, it is sufficient to define the $\eta$ and $\mathit{Re}$, to 
chose $L$ and $\Lambda$, and to calculate from that
$T$ and $\epsilon$. Stability
considerations lead to further constraints for the lattice size, as will be explained in the
following subsection.
\subsection{Stability}
As would be expected, the stability of the simulations over a large number of Monte-Carlo-steps
is a big issue, due to the shock-like solutions of eq. (\ref{dh:eqburg}). Indeed, if certain
restrictions to the lattice parameters are not taken care of rigorously, the simulation terminates 
sooner or later due to divergencies.\\
It is interesting to notice that the occurence of instabilities in our MC-simulations is related to
the (non-trivial) existence of the dissipation scale $\lambda$. We found that to obtain
stable simulations, $\lambda$ has to be resolved on the lattice: $\lambda > \Delta x$. Unstable 
simultations occur otherwise --- we observed that the overall energy of the configurations 
accumulate in the smallest scale $\Delta x$, causing the configuration to separate into two 
sub-configurations, of which one looks as the expected solution to the Hopf-eq., 
while the other grows beyond any limit, eventually breaking the simulation.\\
As long as the dissipation length is resolved, the simulations are stable. Having performed several
millions of MC-steps, no further instabilities occured. If length scales are measured in units of
the system size, this translates into a constraint involving the Reynolds-number:
\begin{equation}
\mathit{Re}<\Lambda L,
\end{equation}
which enforces huge lattices for high $\mathit{Re}$ as $\Lambda \leq 1$.

\subsection{Configurations}
We simulated systems of different sizes, from $(L=4)\times(T=16)$ to larger lattices of the same
viscosity, as $8\times64$ and $16\times256$, and also different viscosities, like $16\times16$ 
or $64\times32$. The Reynolds-number ranged from $\mathit{Re} = 0.01$ to $\mathit{Re} = 100$, not
respecting the above conditions for stability. This may seem surprising, but we could see that
interesting information could also be extracted from the physical sub-configurations. Stable
runs have been obtained for $\mathit{Re} = 1$.\\
Long runs of several millions of MC-steps, on a single node, are realistic for small lattices like
$4\times16$ or $8\times64$ only. For larger lattices, a parallelized version of the
code will be needed.\\
We obtained the following results:
\begin{itemize}
\item{Thermalization and autocorrelation times are very long, up to the order of $10^5$ MC-steps.}
\item{In the stable runs, after thermalization, the typical shock solutions of the Burgers' eq. 
form and can be observed moving and interacting with each other, see fig. (\ref{dh:fig2}).
\begin{figure}
\begin{center}
\includegraphics[width=2.25in]{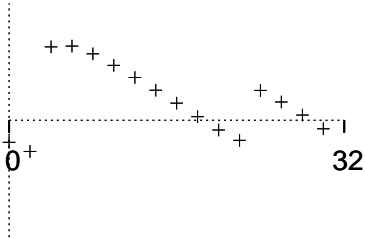}
\caption{Calculated configuration $v(x, t)$ of a $(L=32)\times(T=8)$-lattice; the plot shows a
slize of constant $t$, periodic boundary conditions in $x$. Two shock-like structures are clearly
visible-}
\label{dh:fig2}
\end{center}
\end{figure}
}
\item{In the unstable runs before occurence of the instability, configurations resemble the 
kink-solutions of the Hopf-eq..}
\item{The disctinction between stable and unstable simulations can directly be related to the
existence of a dissipation length scale which is either bigger (stable) or smaller (unstable
simulations) than the lattice spacing.}
\end{itemize}
\section{Summary and Outlook}
We have shown how to perform stable MC-simulations of stochastic partial differential 
eq., like Burgers' or the Hopf-eq.. The lattice versions of the theories can directly be
identified with their continuum counterparts, and, as long as certain constraints on the
lattice size are respected, unlimited numbers of configurations produced. Direct insight
into the structures leading to intermittency and, thus, multiscaling, can be obtained.
Especially, we want to point out that the existence of a dissipation length scale can be
observed.\\
As next steps, complete statistics will be made; especially the scaling exponents of the 
structure functions
have to be compared with the analytic results. Later, we will proceed by analyzing the
incompressible Navier-Stokes-eqs..
\section{Acknowledgements}
We especially want to thank Gernot M\"{u}nster, Westf\"{a}lische Wilhelms-Universit\"{a}t 
M\"{u}nster, for his support in numerous discussions.

\end{document}